\documentclass[aps,twocolumn]{revtex4-1}
\usepackage{amssymb}
\usepackage{amsmath}
\usepackage[dvipsnames]{xcolor}
\usepackage[many]{tcolorbox}
\usepackage{graphicx}
\usepackage{lipsum}
\usepackage{graphicx}

\begin{document}

\title{Polarity of the fermionic condensation in the $p$-wave Kitaev model
on a square lattice}
\author{E. S. Ma}
\author{Z. Song}
\email{songtc@nankai.edu.cn}
\affiliation{School of Physics, Nankai University, Tianjin 300071, China}
\begin{abstract}
In a $p$-wave Kitaev model, the nearest neighbor pairing term results in the
formation of the Bardeen-Cooper-Schrieffer (BCS) pair in the ground state.
In this work, we study the fermionic condensation of real-space pairs in a $%
p $-wave Kitaev model on a square lattice with a uniform phase gradient
pairing term along both directions. The exact solution shows that the ground
state can be expressed in a coherent-state-like form, indicating the
condensation of a collective pairing mode, which is the superposition of
different configurations of pairs in real space. The amplitudes of each
configuration depend not only on the size but also on the orientation of the
pair. We employ three quantities to characterize the ground state in the
thermodynamic limit. (i) A BCS-pair order parameter is introduced to
characterize the phase diagram, consisting of gapful and topological gapless
phases. (ii) The particle-particle correlation length is obtained to reveal
the polarity of the pair condensation. In addition, (iii) a pair-pair
correlator is analytically derived to indicate the possessing of
off-diagonal long-range order. Our work proposes an alternative method for
understanding fermionic condensation.
\end{abstract}

\maketitle

\section*{Introduction}

The concept of fermionic condensate has always received much attention over
a long period of time, from the first discovery of superconductivity in
metal to the first atomic fermionic condensate created by using potassium-40
atoms \cite{Regal2004Observation}. A fermionic condensate is a superfluid
phase formed by fermionic particles at low temperatures. It can describe the
state of electrons in a superconductor or the analogous to fermionic atoms.
For the exact results on the condensation of fermionic pairs in a realistic
model Hamiltonian, we can cast back for much earlier investigations of
excited eta-pairing eigenstates in the Hubbard model for electrons, which
possess off-diagonal long-range order (ODLRO) \cite%
{Yang1962concept,yang1989eta, yang1990so}. Although, a fermionic condensate
is closely related to the Bose--Einstein condensate, the fermion pair has
its own intrinsic structure, for instance, the size and orientation of the
two-fermion dimer. In the Hubbard model, two electrons with opposite spins
can form an on-site pair $c_{\mathbf{r,\uparrow }}^{\dag }c_{\mathbf{%
r,\downarrow }}^{\dag }\left\vert 0\right\rangle $, known as a doublon,
which acts as a hardcore boson. In this context, a many-body eta-pairing
state $\left( \eta ^{+}\right) ^{n}\left\vert 0\right\rangle $\ describes
the condensation of $n$ doublons in a collective mode $\eta ^{+}=\sum_{%
\mathbf{r}}(-1)^{\mathbf{r}}c_{\mathbf{r,\uparrow }}^{\dag }c_{\mathbf{%
r,\downarrow }}^{\dag } $, which is a superposition of single-doublon states
at different lattice sites. On the other hand, an on-site pair is forbidden
for a triplet-pairing mechanism \cite{Ma2023ODLRO,Kitaev}, where the
fundamental building block is the spinless fermion, and then the real-space
pair cannot be regarded as a point particle. Interesting questions are
whether a similar collective pair mode can exist in a spinless fermionic
system and whether condensation will occur in the ground state.

In this paper, we will present some exact results for a $p$-wave Kitaev
model on a square lattice. As well known, the nearest neighbor pairing term
results in the formation of a Bardeen-Cooper-Schrieffer (BCS) pair in the
ground state. In this work, we study the ground state from the perspective
of fermionic condensation of real-space pairs. We consider the $p$-wave
Kitaev model on a square lattice with a uniform phase gradient pairing term
along both directions. The phase gradient provides another dimension in the
phase diagram. We employ three quantities to characterize the ground state
in the thermodynamic limit. (i) A BCS-pair order parameter is introduced to
characterize the phase diagram, consisting of gapful and topological gapless
phases. (ii) The particle-particle correlation length is obtained to reveal
the polarity of the pair condensation. In addition, (iii) a pair-pair
correlator is analytically derived to indicate the existence of ODLRO. Our
work proposes an alternative method for understanding fermionic
condensation. The underlying mechanism is that the ground state can be
expressed in a coherent-state-like form, which is a standard formalism for
the condensation of a collective pairing mode. Such a pair mode is the
superposition of different configurations of pairs in real space, which can
be regarded as an extension of the eta-pairing mode. In contrast to the
uniform amplitude in the eta-pairing mode, the amplitudes of each $p$-wave
pairing configuration depend not only on the size but also on the
orientation of the pair.

The organization of the paper is as follows. We begin Sec. \ref{Model and
phase diagram} by discussing the exact solution of the two-dimensional
Kitaev model under periodic boundary conditions, and we present the phase
diagram. In Sec. \ref{BCS-pair order parameter} the BCS-like order parameter
is introduced to characterize the pairing condition of the ground state, and
we investigate the non-analytical behavior of the parameter. The properties
of the ground state and the form of pairs in coordinate space are discussed
in Sec. \ref{Coherent-state-like ground state}. In Sec. \ref{Correlation
length and polarity} we study the correlation of particle-particle under
ground state, which shows that correlation intension is direction-dependent
in thermodynamic limit. In Sec. \ref{Correlator and ODLRO} we derive the
correlator of two pairs, which indicates that there exists ODLRO in the
ground state. In Sec. \ref{Correlator and ODLRO}, we draw the conclusions.
Some necessary proofs and derivations are given in the appendix.

\section{Model and phase diagram}

\label{Model and phase diagram} We consider the Kitaev model on a square
lattice, which is employed to depict $2$D $p$-wave superconductors. The
Hamiltonian of the tight-binding model on a square $N\times N$ lattice takes
the following form: 
\begin{eqnarray}
H &=&-t\sum_{\mathbf{r,a}}e^{i\phi }c_{\mathbf{r}}^{\dag }c_{\mathbf{r+a}%
}-\Delta \sum_{\mathbf{r,a}}c_{\mathbf{r}}^{\dag }c_{\mathbf{r}+\mathbf{a}%
}^{\dag }  \notag \\
&&+\mathrm{H.c.}+\mu \sum_{\mathbf{r}}\left( 2c_{\mathbf{r}}^{\dag }c_{%
\mathbf{r}}-1\right),  \label{H}
\end{eqnarray}%
where $\mathbf{r=}(x,y)$ are the coordinates of lattice sites and $c_{%
\mathbf{r}}$ are the fermion annihilation operators at site $\mathbf{r}$.
Vectors $\mathbf{a}=(a_{x},a_{y})$ are the lattice vectors in the $x$ and $y$
directions. $t$ is the hopping amplitude between neighboring sites, and the
real number $\Delta $ is the strength of the pair operators. The last term
gives the chemical potential. When the periodic boundary is taken, we define 
$(N+x,y)\rightarrow (x,y)$ and $(x,N+y)\rightarrow (x,y)$. {Here, $\phi $
meets the constraint, $N\phi =2\pi n$\ ($n\in Z$), due to the periodicity of 
$c_{j}$. The value of $\phi $ becomes continuous in the thermodynamic limit.
The phase $\phi $ can be caused by the supercurrent flowing along }the $x$
and $y$ directions{.}\ The Kitaev model is known to have a topological
gapless phase for zero {$\phi $} \cite%
{seradjeh2011unpaired,romito2012manipulating,rontynen2014tuning,dmytruk2019majorana,takasan2022supercurrent}%
. In this work, we only consider the influence of $\phi $ and $\mu $ on the
ground state for the Hamiltonian with a fixed $\Delta >0$.

Imposing periodic boundary conditions in both directions, the Hamiltonian
can be exactly diagonalized. Applying the Fourier transformation

\begin{eqnarray}
c_{\mathbf{r}} &=&\frac{1}{N}\sum_{\mathbf{k}}e^{i\mathbf{k}\cdot \mathbf{r}%
}c_{\mathbf{k}}, \\
k_{x},k_{y} &=&\frac{2\pi m}{N},m\in Z.  \notag
\end{eqnarray}%
The Hamiltonian can be written in the form

\begin{equation}
H=2\sum_{\mathbf{k}\in A}[\left( 
\begin{array}{cc}
c_{\mathbf{k}}^{\dag } & c_{-\mathbf{k}}%
\end{array}%
\right) h_{k}\left( 
\begin{array}{c}
c_{\mathbf{k}} \\ 
c_{-\mathbf{k}}^{\dag }%
\end{array}%
\right) -2t\cos k_{+}\cos \left( \phi -k_{-}\right) ],
\end{equation}%
where the domain of the summation is $A=\left\{ \mathbf{k,|}\sin k_{x}-\sin
k_{y}<0\right\} $ and is illustrated in Fig. 1. The choice of $A$ is
multiple and ensures the simplicity of the form of the ground state. Here,
the matrix $h_{\mathbf{k}}$ can be expressed as

\begin{figure}[tbh]
\centering
\includegraphics[width=0.5\textwidth]{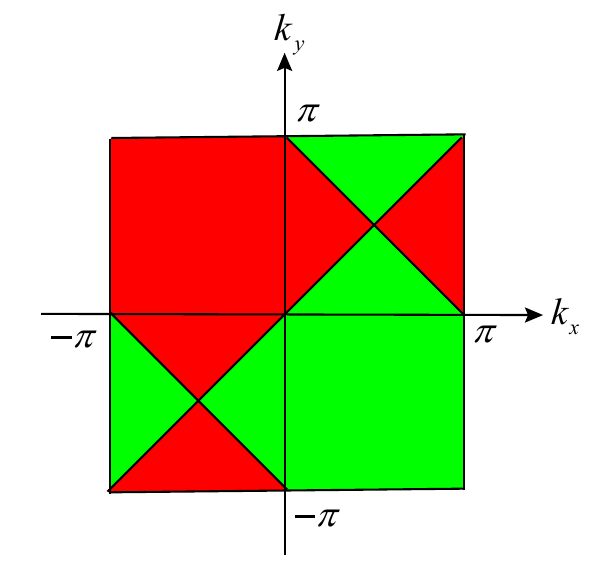}
\caption{The schematic diagram of the Brillouin zone in momentum space. $%
\mathbf{k}=\left( k_{x},k_{y}\right) $ and $k_{x},k_{y}\in \left( -\protect%
\pi ,\protect\pi \right] ,$the whole region can be divided into two parts
according to the sign of $\left( \sin k_{x}-\sin k_{y}\right) $. $\mathbf{k}$
in the red region satisfies $\sin k_{x}-\sin k_{y}<0,$ which is called A in
this paper. }
\label{fig1}
\end{figure}
\begin{eqnarray}
&&h_{\mathbf{k}}=\sum_{i=0}^{3}B_{i}\sigma _{i} \\
&=&\left( 
\begin{array}{cc}
\mu -2t\cos k_{+}\cos \left( \phi +k_{-}\right) & -2i\Delta \cos k_{+}\sin
k_{-} \\ 
2i\Delta \cos k_{+}\sin k_{-} & 2t\cos k_{+}\cos \left( \phi -k_{-}\right)
-\mu%
\end{array}%
\right) ,  \notag
\end{eqnarray}%
with modified wave vectors $k_{\pm }=\left( k_{x}\pm k_{y}\right) /2$ for
the sake of simplicity, where $\left\{ \sigma _{i}\right\} $\ are Pauli
matrices in the form

\begin{eqnarray}
\sigma _{0} &=&\left( 
\begin{array}{cc}
1 & 0 \\ 
0 & 1%
\end{array}%
\right) ,\sigma _{1}=\left( 
\begin{array}{cc}
0 & 1 \\ 
1 & 0%
\end{array}%
\right) ,  \notag \\
\sigma _{2} &=&\left( 
\begin{array}{cc}
0 & -i \\ 
i & 0%
\end{array}%
\right) ,\sigma _{3}=\left( 
\begin{array}{cc}
1 & 0 \\ 
0 & -1%
\end{array}%
\right) .
\end{eqnarray}%
Here, the energy basis $B_{0}=$ $2t\sin \phi \cos k_{+}\sin k_{-}$\ and
components of the auxiliary field $\mathbf{B}\left( \mathbf{k}\right) =$ $%
(B_{1},B_{2},B_{3})$\ are%
\begin{equation}
\left\{ 
\begin{array}{l}
B_{1}=0, \\ 
B_{2}=2\Delta \cos k_{+}\sin k_{-}, \\ 
B_{3}=\mu -2t\cos \phi \cos k_{+}\cos k_{-}.%
\end{array}%
\right.
\end{equation}%
The Hamiltonian can be diagonalized in the form

\begin{equation}
H=\sum_{\mathbf{k\in }\mathrm{BZ}}[2\left( \varepsilon _{\mathbf{k}%
}+B_{0}\right) \left( \gamma _{\mathbf{k}}^{\dag }\gamma _{\mathbf{k}}-\frac{%
1}{2}\right) +B_{0}],
\end{equation}%
by introducing the Bogoliubov operator

\begin{equation}
\gamma _{\mathbf{k}}=\cos \frac{\theta _{k}}{2}c_{\mathbf{k}}+\sin \frac{%
\theta _{k}}{2}e^{-i\varphi }c_{-\mathbf{k}}^{\dag },
\end{equation}%
which satisfies the commutation relations of the fermion within region$\ A$ 
\begin{equation}
\left\{ \gamma _{\mathbf{k}},\gamma _{\mathbf{q}}^{\dagger }\right\} =\delta
_{\mathbf{k},\mathbf{q}},\text{ }\left\{ \gamma _{\mathbf{k}},\gamma _{%
\mathbf{q}}\right\} =\left\{ \gamma _{\mathbf{k}}^{\dagger },\gamma _{%
\mathbf{q}}^{\dagger }\right\} =0.
\end{equation}%
Here, the reduced dispersion relation is 
\begin{equation}
\varepsilon _{\mathbf{k}}=\sqrt{B_{2}^{2}+B_{3}^{2}},
\end{equation}%
where $\theta _{\mathbf{k}}$\ and $\varphi $ are determined by

\begin{eqnarray}
\tan \theta _{\mathbf{k}} &=&\frac{2\Delta \left\vert \cos k_{+}\sin
k_{-}\right\vert }{\mu -2t\cos \phi \cos k_{+}\cos k_{-}}, \\
\sin \varphi &=&\mathrm{sign}\left( \cos k_{+}\sin k_{-}\right) .
\end{eqnarray}%
If $\Delta >t>0$, the ground state can be constructed as%
\begin{equation}
\left\vert \text{G}\left( \mu ,\phi \right) \right\rangle =\prod_{\mathbf{k}%
}\gamma _{\mathbf{k}}\left\vert 0\right\rangle ,
\end{equation}%
with the ground state energy%
\begin{equation}
E_{\mathrm{g}}=-\sum_{\mathbf{k}}\varepsilon _{\mathbf{k}},
\end{equation}%
where $\left\vert 0\right\rangle $ is the vacuum state of $c_{\mathbf{k}}$,
satisfying $c_{\mathbf{k}}\left\vert 0\right\rangle =0$ for all $\mathbf{k}$%
. We note that the gapless ground state appears when $\varepsilon _{\mathbf{k%
}}$ has a zero point or band touching point of a single $\gamma _{\mathbf{k}%
} $-particle spectrum.

The band degenerate point $\mathbf{k}_{0}=(k_{0+},k_{0-})$ is determined by%
\begin{equation}
\left\{ 
\begin{array}{l}
\cos k_{+}\sin k_{-}=0, \\ 
\mu -2t\cos \phi \cos k_{+}\cos k_{-}=0.%
\end{array}%
\right.
\end{equation}%
There exist two phases, gapless and gapful, with the boundary at curve $%
\left\vert \mu /(2t)\right\vert =\cos \phi $, i.e.,

\begin{equation}
\left\{ 
\begin{array}{c}
\left\vert \mu \right\vert <\left\vert 2t\cos \phi \right\vert ,\text{ }%
\mathrm{gapless}, \\ 
\left\vert \mu \right\vert >\left\vert 2t\cos \phi \right\vert ,\text{ }%
\mathrm{gapful}.%
\end{array}%
\right.
\end{equation}%
In the gapless phase, the degenerate points are isolated and can be
classified as topologically trivial and nontrivial using topological numbers
or invariants \cite{Sun2012topological,wang2017gapless,JMH}. 
\begin{figure}[tbh]
\centering
\includegraphics[width=0.5\textwidth]{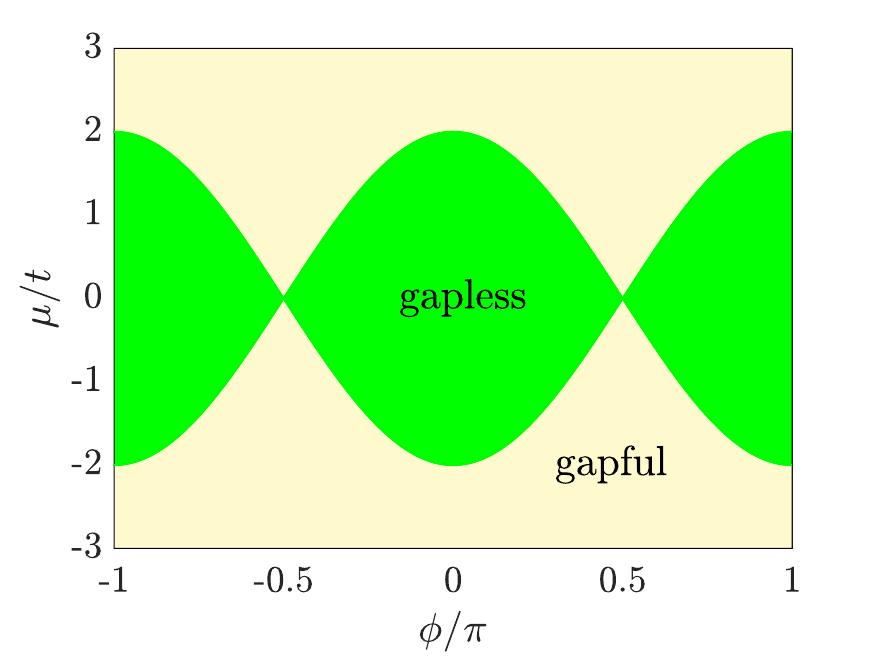}
\caption{Phase diagram of the Hamiltonian in Eq. (\protect\ref{H}) on the
parameter $\protect\phi -\protect\mu /t$ plane. Different color regions
represent different phases that are distinguished by whether there exists an
energy gap between two eigenstates with opposite $\mathbf{k}$. The
boundaries are determined by the equation $\left\vert \protect\mu /\left(
2t\right) \right\vert =\left\vert \cos \protect\phi \right\vert $. The green
region is a nontrivial phase, in which the topological number is nonzero and
the derivative of the eigenenergy is discontinuous. }
\label{fig2}
\end{figure}

\begin{figure*}[tbh]
\centering
\includegraphics[width=1.0\textwidth]{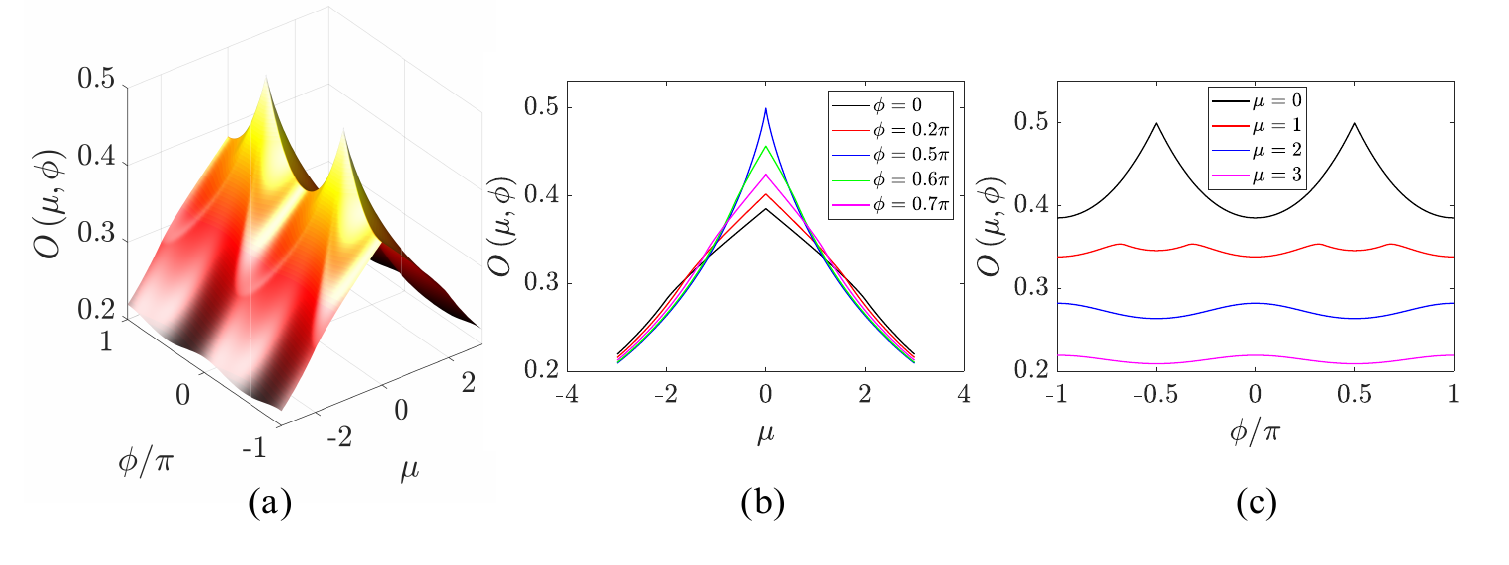}
\caption{(a) Color contour plots of the numerical results of order parameter 
$O_{\text{g}}\left( \protect\mu ,\protect\phi \right) $ defined in Eq. (%
\protect\ref{O}). (b) Plots of $O_{\text{g}}\left( \protect\mu ,\protect\phi %
_{0}\right) $ for several representative values of $\protect\phi _{0}$. (c)
Plots of $O_{\text{g}}\left( \protect\mu _{0},\protect\phi \right) $ for
several different $\protect\mu _{0}$. The parameters are $N=2000$, $t=1$ and 
$\Delta =2$. (a) shows that the order parameter reaches its maxima 1/2 at
the triple critical points $\left( \protect\mu ,\protect\phi \right) =\left(
0,\pm \protect\pi /2\right)$. (b) and (c) indicate that $O$ is smooth if the
parameters in the gapful region but has nonanalytic points in the gapless
region.}
\label{fig3}
\end{figure*}
\begin{figure*}[tbh]
\centering
\includegraphics[width=1.0\textwidth]{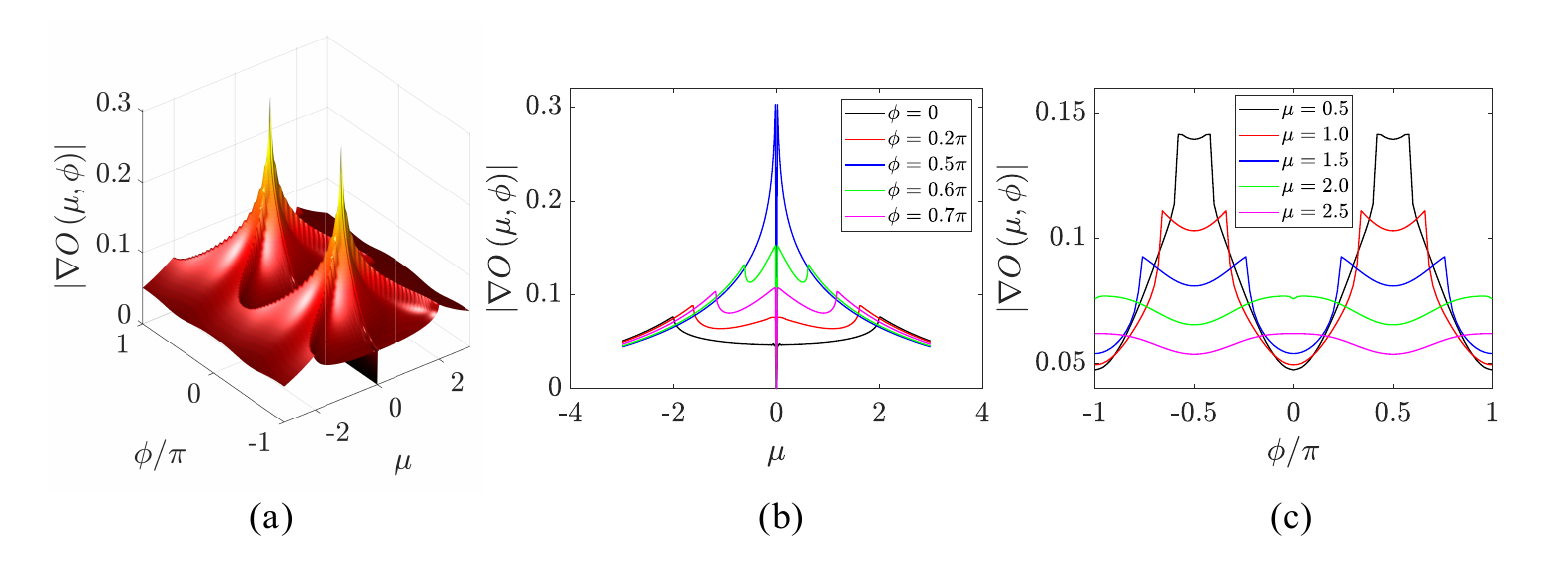}
\caption{(a) Color contour plots of numerical results of the absolute value
of the gradient of the order parameter $\left\vert \protect\nabla O_{\text{g}%
}\left( \protect\mu ,\protect\phi \right) \right\vert $ defined in Eq. (%
\protect\ref{dO}). (b) Plots of $\left\vert \protect\nabla O_{\text{g}%
}\left( \protect\mu ,\protect\phi _{0}\right) \right\vert $ for several
representative values of $\protect\phi _{0}$. (c) Plots of $\left\vert 
\protect\nabla O_{ \text{g}}\left( \protect\mu _{0},\protect\phi \right)
\right\vert$ for several different $\protect\mu _{0}$. The parameters are $%
N=2000$, $t=1$ and $\Delta =2$. We can see that there are boundaries between
different regions in (a), which are phase boundaries in Fig. \protect\ref%
{fig2}, and $\left( \protect\mu ,\protect\phi \right) =\left( 0,\pm \protect%
\pi /2\right)$ are nonanalytic points. (b) shows that phase boundaries are
determined by the equations $\left\vert \protect\mu \right\vert =\left\vert
2t\cos \protect\phi \right\vert$ and (c) indicates $\left\vert \protect%
\nabla O_{\text{g} }\left( \protect\mu ,\protect\phi\right) \right\vert $
has nonanalytic points at the boundaries if the parameters are in the
gapless region. }
\label{fig4}
\end{figure*}

\section{BCS-pair order parameter}

\label{BCS-pair order parameter}

In this section, we investigate the phase diagram from another point of
view. We introduce a BCS-pair order parameter, which is a quantity to
describe the rate of the transition from free fermions to BCS pairs. It is
related to a set of pseudospin operators, which is defined as%
\begin{eqnarray}
s_{\mathbf{k}}^{+} &=&\left( s_{\mathbf{k}}^{-}\right) ^{\dag }=c_{-\mathbf{k%
}}^{\dag }c_{\mathbf{k}}^{\dag },  \notag \\
s_{\mathbf{k}}^{z} &=&\frac{1}{2}\left( c_{-\mathbf{k}}^{\dag }c_{-\mathbf{k}%
}+c_{\mathbf{k}}^{\dag }c_{\mathbf{k}}-1\right) ,
\end{eqnarray}%
satisfying the Lie algebra commutation relations

\begin{equation}
\left[ s_{\mathbf{k}}^{+},s_{\mathbf{k}}^{-}\right] =2s_{\mathbf{k}}^{z},%
\left[ s_{\mathbf{k}}^{z},s_{\mathbf{k}}^{\pm }\right] =\pm s_{\mathbf{k}%
}^{\pm }.
\end{equation}%
Obviously, $s_{\mathbf{k}}^{\pm }$\ is the BCS-pair creation or annihilation
operator in the $\mathbf{k}$-channel, while $s_{\mathbf{k}}^{z}$ relates to
the pair number.

We introduce an order parameter

\begin{equation}
O_{\text{g}}=\frac{2}{N^{2}}\sum_{\mathbf{k\in }A}\left\vert \left\langle 
\text{G}\left( \mu ,\phi \right) \right\vert s_{\mathbf{k}}^{+}\left\vert 
\text{G}\left( \mu ,\phi \right) \right\rangle \right\vert ,  \label{O}
\end{equation}%
which is the sum of the expectation absolute value of $c_{-\mathbf{k}}^{\dag
}c_{\mathbf{k}}^{\dag }$\ for the ground state, characterizing the pairing
progress. The exact expression of the ground state results in%
\begin{equation}
O_{\text{g}}=\frac{\Delta }{N^{2}}\sum_{\mathbf{k\in }A}\frac{\sin
k_{y}-\sin k_{x}}{\varepsilon _{\mathbf{k}}},
\end{equation}%
which indicates that $O_{\text{g}}$\ is a quantity intimately related to the
spectrum $\varepsilon _{\mathbf{k}}$. It is expected that the analytic
behavior of $O_{\text{g}}$ should demonstrate the phase diagram. To this
end, we plot $O_{\text{g}}$ and the corresponding derivatives\ 
\begin{eqnarray}
\frac{\partial O_{\text{g}}}{\partial \mu } &=&\frac{1}{N^{2}}\sum_{\mathbf{%
k\in }A}\frac{B_{2}B_{3}}{\varepsilon _{\mathbf{k}}^{3}}, \\
\frac{\partial O_{\text{g}}}{\partial \phi } &=&\frac{1}{N^{2}}\sum_{\mathbf{%
k\in }A}\frac{J\sin \phi \left( \cos k_{x}+\cos k_{y}\right) B_{2}B_{3}}{%
\varepsilon _{\mathbf{k}}^{3}},
\end{eqnarray}%
and%
\begin{equation}
\left\vert \triangledown O_{\text{g}}\left( \mu ,\phi \right) \right\vert =%
\sqrt{\left( \frac{\partial O_{\text{g}}}{\partial \mu }\right) ^{2}+\left( 
\frac{\partial O_{\text{g}}}{\partial \phi }\right) ^{2}},  \label{dO}
\end{equation}%
in Fig. \ref{fig4}. We find that $O_{\text{g}}$\ attains its maximum at the
point $\left( 0,\pi /2\right) $, and there is an evident jump at the quantum
phase boundary.

\section{Coherent-state-like ground state}

\label{Coherent-state-like ground state}

We note that the expression of the order parameter $O_{\text{g}}$\ in Eq. (%
\ref{O}) can also be written in the form 
\begin{equation}
O_{\text{g}}=\frac{2}{N^{2}}\left\vert \left\langle \text{G}\left( \mu ,\phi
\right) \right\vert s^{+}\left\vert \text{G}\left( \mu ,\phi \right)
\right\rangle \right\vert ,
\end{equation}%
due to the fact that $\sin k_{y}-\sin k_{x}$\ is always positive within 
region$\ A$. Here, a set of operators%
\begin{equation}
s^{\pm }=\sum_{\mathbf{k\in }A}s_{\mathbf{k}}^{\pm },s^{z}=\sum_{\mathbf{%
k\in }A}s_{\mathbf{k}}^{z},  \label{s}
\end{equation}%
are still pseudo spin operators, satisfying the Lie algebra commutation
relations 
\begin{equation}
\left[ s^{+},s^{-}\right] =2s^{z},\left[ s^{z},s^{\pm }\right] =\pm s^{\pm }.
\end{equation}%
The finite value of $O_{\text{g}}$\ indicates that the ground state $%
\left\vert \text{G}\left( \mu ,\phi \right) \right\rangle $\ is probably the
superposition of the set of states 
\begin{equation}
\left\vert \psi _{n}\right\rangle =\frac{1}{\Omega _{n}}\left( s^{+}\right)
^{n}\left\vert 0\right\rangle ,n\in \left[ 0,N^{2}/2\right] ,
\end{equation}%
in some cases. In fact, the ground state can be expressed explicitly in
terms of $\theta _{\mathbf{k}}$ 
\begin{equation}
\left\vert \text{G}\left( \mu ,\phi \right) \right\rangle =\prod\limits_{%
\mathbf{k}\in A}\left( i\cos \frac{\theta _{\mathbf{k}}}{2}+\sin \frac{%
\theta _{\mathbf{k}}}{2}c_{\mathbf{k}}^{\dag }c_{-\mathbf{k}}^{\dag }\right)
\left\vert 0\right\rangle _{\mathbf{k}}\left\vert 0\right\rangle _{-\mathbf{k%
}},
\end{equation}%
which can be expressed as a tensor product state of a pair in real space. In
general, for an arbitrary tensor product state in the form

\begin{equation}
\left\vert \phi (\beta )\right\rangle =\prod_{l=1}^{m}\left[ -ie^{i\alpha
l}\sin \left( \beta /2\right) s_{l}^{+}+\cos \left( \beta /2\right) \right]
\left\vert \downarrow \right\rangle _{l},
\end{equation}%
where the set of spin (or hardcore boson) operators $\left\{
s_{l}^{+}\right\} $\ of number $m$, has actions $s_{l}^{+}\left\vert
\downarrow \right\rangle _{l}=\left\vert \uparrow \right\rangle _{l}$, $%
s_{l}^{+}\left\vert \uparrow \right\rangle _{l}=0$,\ and $\left(
s_{l}^{+}\right) ^{\dag }\left\vert \downarrow \right\rangle _{l}=0$. The
parameters $\alpha $\ and $\beta $\ are two arbitrary angles to determine
the profile of the state. Direct derivation shows that it can also be
written in the form 
\begin{equation}
\left\vert \phi (\beta )\right\rangle =\sum_{n=1}^{m}d_{n}\left\vert \psi
_{n}\right\rangle ,
\end{equation}%
where a set of states $\left\{ \left\vert \psi _{n}\right\rangle ,n\in \left[
0,m\right] \right\} $\ can be constructed as\ 

\begin{equation}
\left\vert \psi _{n}\right\rangle =\frac{1}{\left( n!\right) \sqrt{C_{m}^{n}}%
}\left( \sum_{l=1}^{m}e^{i\alpha l}s_{l}^{+}\right)
^{n}\prod\limits_{l=1}^{m}\left\vert \downarrow \right\rangle _{l},
\end{equation}%
$\left\vert \psi _{n}\right\rangle $ is defined as 
\begin{equation}
d_{n}=\sqrt{C_{m}^{n}}\left( -i\right) ^{n}\sin ^{n}\left( \beta /2\right)
\cos ^{\left( m-n\right) }\left( \beta /2\right) .
\end{equation}

Note that $\left\vert \psi _{n}\right\rangle $\ is in the form of a
coherent-state-like, which always exhibits the feature of condensation in
the $\mathbf{k}$ space\ and ODLRD in the real space. One can apply the above
result to span a part of the product state of the ground state

\begin{equation}
\prod\limits_{\mathbf{k}\in C}\left( i\cos \frac{\theta _{\mathbf{k}}}{2}%
+\sin \frac{\theta _{\mathbf{k}}}{2}c_{\mathbf{k}}^{\dag }c_{-\mathbf{k}%
}^{\dag }\right) \left\vert 0\right\rangle _{\mathbf{k}}\left\vert
0\right\rangle _{-\mathbf{k}},
\end{equation}%
with $A\supseteq C$, in which $\theta _{\mathbf{k}}$\ is $\mathbf{k}$
independent. For instance, taking $\mu =0$, we have%
\begin{equation}
\tan \theta _{\mathbf{k}}=\frac{\Delta }{t\cos \phi }\tan k_{-},
\end{equation}%
which ensures the constancy of $\theta _{\mathbf{k}}$ if the set $C$\
corresponds to the line with fixed $k_{-}$\ in the BZ. In this work, we
focus on an extreme case with $\phi =\pi /2$, which ensures $A=C$, and then 
\begin{eqnarray}
\left\vert \text{G}\left( 0,\pi /2\right) \right\rangle &=&\prod\limits_{%
\mathbf{k\in }A}\frac{1-ic_{\mathbf{k}}^{\dag }c_{-\mathbf{k}}^{\dag }}{%
\sqrt{2}}\left\vert 0\right\rangle _{\mathbf{k}}\left\vert 0\right\rangle _{-%
\mathbf{k}}  \notag \\
&=&\sum_{n=0}^{N^{2}/2}\frac{i^{n}2^{-N^{2}/4}}{n!}\left( s^{+}\right)
^{n}\left\vert 0\right\rangle .
\end{eqnarray}%
It is obvious that $\left\vert \text{G}\left( 0,\pi /2\right) \right\rangle $%
\ represents the condensates of a collective BCS-like pair, which is a
superposition of all possible pairs $c_{-\mathbf{k}}^{\dag }c_{\mathbf{k}%
}^{\dag }$ in $\mathbf{k}$ space. Importantly, it can also be regarded as
the condensate of a collective pair in real space due to the following fact.
In the thermodynamic limit, a direct derivation from the Appendix shows

\begin{equation}
s^{+}=\frac{2i}{\pi ^{2}}\sum_{\mathbf{\mathbf{r,r}}^{\prime }}\frac{\sin %
\left[ \pi \left( \Delta x+\Delta y\right) /2\right] }{\left( \Delta
y\right) ^{2}-\left( \Delta x\right) ^{2}}c_{\mathbf{r}}^{\dag }c_{\mathbf{r}%
^{\prime }}^{\dag },  \label{collective operator}
\end{equation}%
where the displacement $\mathbf{r^{\prime }-\mathbf{r}}$\textbf{\textbf{\ }}%
has two components\textbf{\textbf{\ }}$(\Delta x,\Delta y)$. It is a
superposition of all possible pairing configurations with the $(\Delta
x,\Delta y)$-dependent amplitudes and is then referred to as the collective
pair operator. Notably, the amplitudes are zero when $\Delta x+\Delta y$\ is
even. The nonzero amplitudes decay in power law with exponent $2$ in the $x$
and $y$ directions and have strong polarity. In the following section, we
study this feature by means of a correlation function.

\section{Correlation length and polarity}

\label{Correlation length and polarity}

The exact expression for the ground state in Eq. (\ref{collective operator})
allows us to investigate the details of the fermion condensates. In this
section, we calculate the particle-particle correlation function between two
sites $\left( 0,0\right) $ and $\left( x,y\right) $

\begin{equation}
g\left( r,\theta \right) =\left\vert \left\langle \text{G}\left( 0,\pi
/2\right) \right\vert c_{\left( 0,0\right) }^{\dag }c_{\left( x,y\right)
}^{\dag }\left\vert \text{G}\left( 0,\pi /2\right) \right\rangle \right\vert
,  \label{g}
\end{equation}%
which represents the correlation intension of the pairs with the distance $r=%
\sqrt{x^{2}+y^{2}}$ in the $\theta $ direction. The coherent-state-like
ground state yields 
\begin{equation}
g\left( r,\theta \right) =\frac{1}{N^{2}}\left\vert \sum_{\mathbf{k\in }%
A}\sin \left( k_{x}x+k_{y}y\right) \right\vert ,
\end{equation}%
which becomes%
\begin{equation}
g\left( r,\theta \right) =\frac{1}{4\pi ^{2}}\left\vert \int_{\mathbf{k\in }%
A}\sin \left( k_{x}x+k_{y}y\right) \text{\textrm{d}}k_{x}\mathrm{d}%
k_{y}\right\vert ,
\end{equation}%
in the thermodynamic limit. A straightforward derivation leads to the
following result 
\begin{eqnarray}
g\left( r,\theta \right) &=&\frac{2}{\pi ^{2}\left\vert
y^{2}-x^{2}\right\vert },\left( x+y=\mathrm{odd}\right)  \notag \\
&=&\frac{2}{\pi ^{2}r^{2}\left\vert \cos \left( 2\theta \right) \right\vert }%
,
\end{eqnarray}%
power law decay, indicating an infinite correlation length. This agrees with
the fact that $\left( \mu ,\phi \right) =$ $\left( 0,\pi /2\right) $ is the
critical point with gapless spectrum in the phase diagram. Nevertheless, the
relative correlation intension of the pairs in different directions, i.e.,
the exponent is polarized and can be obtained from

\begin{equation}
\alpha \left( \infty ,\theta \right) =-\lim_{r\longrightarrow \infty }\frac{%
\ln g\left( r,\theta \right) +\ln \left( \pi ^{2}/2\right) }{\ln r}.
\label{alpha}
\end{equation}%
We have 
\begin{eqnarray}
&&\alpha \left( \infty ,0\right) =\alpha \left( \infty ,\pi /2\right) =2, 
\notag \\
&&\alpha \left( \infty ,\pi /4\right) =1.
\end{eqnarray}%
We plot the exponent in Eq. (\ref{alpha}) in Fig. \ref{fig4}. We find that $%
\alpha $\ is approximately $2$ but only experiences a sharp dip at $\theta
=\pi /4$.

\begin{figure}[tbh]
\centering
\includegraphics[width=0.5\textwidth]{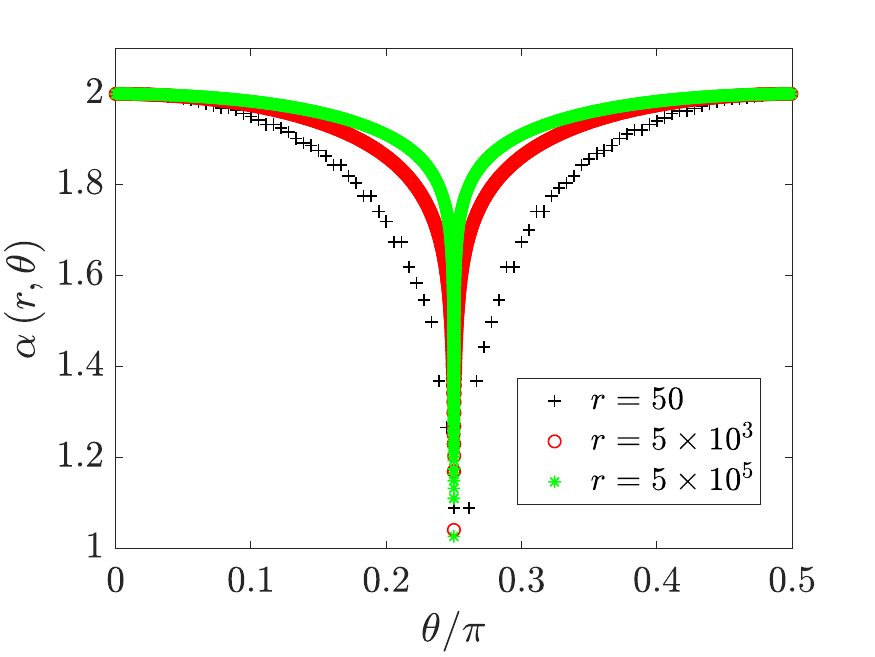}
\caption{The numerical results of polarity defined in Eq. (\protect\ref%
{alpha}). The results show that $\protect\alpha$ is 2 with $\protect\theta =0
$ or $\protect\theta =\protect\pi /2$ but approaches 1 with $\protect\theta =%
\protect\pi /4$ as $r$ increases, and there is a drastic change in the
vicinity of $\protect\pi /4$. This indicates that the correlation intension
of pairs is different in different directions. }
\label{fig5}
\end{figure}

\section{Correlator and ODLRO}

\label{Correlator and ODLRO}

It is clear that state $\left\vert G\right\rangle $\ is a coherent state of
zero-momentum-pair condensation. In the following, we will show that such a
state possesses ODLRO. To this end, we introduce the correlator 
\begin{equation}
C_{\mathbf{RR}^{\prime }}=\left\vert \left\langle \text{G}\left( 0,\pi
/2\right) \right\vert \left( c_{\mathbf{r}}c_{\mathbf{r}+\Delta \mathbf{r}%
}\right) ^{\dag }c_{\mathbf{r}^{\prime }}c_{\mathbf{r}^{\prime }+\Delta 
\mathbf{r}^{\prime }}\left\vert \text{G}\left( 0,\pi /2\right) \right\rangle
\right\vert ,  \label{C}
\end{equation}%
where $\mathbf{R=r}+\Delta \mathbf{r}$\ and $\mathbf{R}^{\prime }\mathbf{=r}%
^{\prime }+\Delta \mathbf{r}^{\prime }$. It {measures the correlation of two
pairs and characterizes the condensation of pairs. }Under the condition $%
\left\vert \mathbf{r}^{\prime }\mathbf{-r}\right\vert \gg \left\vert \Delta 
\mathbf{r}\right\vert ,\left\vert \Delta \mathbf{r}^{\prime }\right\vert $,
two pairs can be regarded as bosons located at $\mathbf{r}$\ and $\mathbf{r}$%
, i.e., $\left( c_{\mathbf{r}}c_{\mathbf{r}+\Delta \mathbf{r}}\right) ^{\dag
}\rightarrow b_{\mathbf{r}}^{\dag }$ and $c_{\mathbf{r}^{\prime }}c_{\mathbf{%
r}^{\prime }+\Delta \mathbf{r}^{\prime }}\longrightarrow b_{\mathbf{r}%
^{\prime }}$. Then, for a boson condensate state 
\begin{equation}
\left\vert b_{\mathbf{k}},n\right\rangle =\frac{1}{\sqrt{n!}}\left( \frac{1}{%
N}\sum_{\mathbf{r}}e^{i\mathbf{k\cdot r}}b_{\mathbf{r}}^{\dag }\right)
^{n}\left\vert 0\right\rangle ,
\end{equation}%
we have%
\begin{equation}
\left\vert \left\langle b_{\mathbf{k}},n\right\vert b_{\mathbf{r}}^{\dag }b_{%
\mathbf{r}^{\prime }}\left\vert b_{\mathbf{k}},n\right\rangle \right\vert
\sim \frac{n}{N^{2}},
\end{equation}%
which identifies the ODLRO{\ of the state }$\left\vert b,n\right\rangle ${.
In parallel, one can employ the correlator }$C_{\mathbf{RR}^{\prime }}${\ to
identify the condensation of fermion pairs. In the thermodynamic limit, the
direct derivation of }$C_{\mathbf{RR}^{\prime }}${\ in the Appendix shows
that }%
\begin{equation}
C_{\mathbf{RR}^{\prime }}=g\left( \Delta r,\theta \right) g\left( \Delta
r^{\prime },\theta ^{\prime }\right) ,  \label{CRR'}
\end{equation}%
for $\left\vert \mathbf{r}^{\prime }\mathbf{-r}\right\vert \gg \left\vert
\Delta \mathbf{r}\right\vert ,\left\vert \Delta \mathbf{r}^{\prime
}\right\vert $. {Then, we conclude that the ground state at the triple
critical point possesses the exact ODLRO. In comparison with the real
bosons, the values of }$C_{\mathbf{RR}^{\prime }}${\ depend on the parity of 
}$\left( \Delta r,\Delta r^{\prime }\right) $ and the orientation of $\left(
\theta ,\theta ^{\prime }\right) $, since the fermion pair is not a point
particle and has its own intrinsic structure. In this sense, {the correlator 
}$C_{\mathbf{RR}^{\prime }}${\ can identify not only condensation but also
the pairing mechanism.}\ 

\section{Summary}

\label{Summary} In summary, we have provided a concrete example to
demonstrate the fermonic condensate in a spinless system, which is different
from the eta-pairing mechanism in a spinful system, such as the Hubbard
model. We have shown that the ground state can be expressed in two forms:
(i) a BCS-like state, which consists of two fermions with opposite momentum,
and (ii) a coherent-state-like form, which is a standard formalism for the
condensation of a collective pairing mode. In this sense, such a pair mode
can be regarded as an extension of the eta-pairing mode. We also employed
three quantities to characterize the ground state in the thermodynamic
limit. The BCS-pair order parameter can be used to identify the phase
diagram. The two-operator correlation function reveals the polarity of the
pair condensation. The four-operator correlator indicates the ODLRO as
expected. Taken together, these results unveil the mechanism\ of fermionic
condensate in the p-wave superconducting state.

\section*{Acknowledgements}

We acknowledge the support of the CNSF (Grant No. 11874225).

\section{Appendix}

\appendix\setcounter{equation}{0} \renewcommand{\theequation}{A%
\arabic{equation}} \setcounter{figure}{0} \renewcommand{\thefigure}{A	%
\arabic{figure}} In this Appendix, we present a derivation of the expression
of $s^{+}$\ in coordinate space, Eq. (\ref{s}) in the main text, based on
which we obtained two correlation functions in Eq. (\ref{g}) and Eq. (\ref{C}%
), respectively.

Applying the Fourier transformation 
\begin{equation}
c_{\mathbf{k}}=\frac{1}{N}\sum_{\mathbf{r}}e^{-i\mathbf{k}\cdot \mathbf{r}%
}c_{\mathbf{r}},
\end{equation}%
we have 
\begin{eqnarray}
s^{+} &=&\sum_{\mathbf{k\in }A}c_{-\mathbf{k}}^{\dag }c_{\mathbf{k}}^{\dag }=%
\frac{1}{N^{2}}\sum_{\mathbf{k\in }A}\sum_{\mathbf{r,r}^{\prime }}e^{i%
\mathbf{k}\cdot \left( \mathbf{r}^{\prime }-\mathbf{r}\right) }c_{\mathbf{r}%
}^{\dag }c_{\mathbf{r}^{\prime }}^{\dag }  \notag \\
&=&\frac{i}{N^{2}}\sum_{\mathbf{k\in }A}\sum_{\mathbf{r,r}^{\prime }}\sin %
\left[ \mathbf{k}\cdot \left( \mathbf{r}^{\prime }-\mathbf{r}\right) \right]
c_{\mathbf{r}}^{\dag }c_{\mathbf{r}^{\prime }}^{\dag }  \notag \\
&=&i\sum_{\mathbf{r,r}^{\prime }}\Gamma \left( \mathbf{r}^{\prime }-\mathbf{r%
}\right) c_{\mathbf{r}}^{\dag }c_{\mathbf{r}^{\prime }}^{\dag },
\end{eqnarray}%
where $\Gamma \left( \mathbf{r}^{\prime }-\mathbf{r}\right) $\ is only the
function of $\mathbf{r^{\prime }-\mathbf{r=}}(\Delta x,\Delta y)$ 
\begin{equation}
\Gamma \left( \mathbf{r}^{\prime }-\mathbf{r}\right) =\frac{1}{N^{2}}\sum_{%
\mathbf{k\in }A}\sin \left[ \mathbf{k}\cdot \left( \mathbf{r}^{\prime }-%
\mathbf{r}\right) \right] .
\end{equation}%
In the thermodynamic limit, it can be integrated as the function of $(\Delta
x,\Delta y)$,%
\begin{eqnarray}
\Gamma \left( \mathbf{r}^{\prime }-\mathbf{r}\right) &=&\frac{1}{4\pi ^{2}}%
\int_{\mathbf{k\in }A}\sin \left[ \mathbf{k}\cdot \left( \mathbf{r}^{\prime
}-\mathbf{r}\right) \right] \text{d}^{2}\mathbf{k}  \notag \\
&=&\frac{2}{\pi ^{2}}\frac{\sin \left[ \pi \left( \Delta x+\Delta y\right) /2%
\right] }{\left( \Delta y\right) ^{2}-\left( \Delta x\right) ^{2}},
\end{eqnarray}%
and then we have%
\begin{equation}
s^{+}=\frac{2i}{\pi ^{2}}\sum_{\mathbf{\mathbf{r,r}}^{\prime }}\frac{\sin %
\left[ \pi \left( \Delta x+\Delta y\right) /2\right] }{\left( \Delta
y\right) ^{2}-\left( \Delta x\right) ^{2}}c_{\mathbf{r}}^{\dag }c_{\mathbf{r}%
^{\prime }}^{\dag }.
\end{equation}

On the other hand, inversely, we have

\begin{gather}
c_{\mathbf{r}}c_{\mathbf{r}^{\prime }}=\frac{1}{N^{2}}\sum_{\mathbf{k,k}%
^{\prime }}e^{i\mathbf{k\cdot r}}e^{i\mathbf{k}^{\prime }\cdot \mathbf{r}%
^{\prime }}c_{\mathbf{k}}c_{\mathbf{k}^{\prime }}=\frac{1}{N^{2}}\times 
\notag \\
\sum_{\mathbf{k,k}^{\prime }\mathbf{\in }A,\left( \mathbf{k}^{\prime }\neq 
\mathbf{k}\right) }\{[e^{i\mathbf{k\cdot r}}e^{-i\mathbf{k}^{\prime }\cdot 
\mathbf{r}^{\prime }}-(\mathbf{k\leftrightarrows -k}^{\prime })]c_{\mathbf{k}%
}c_{-\mathbf{k}^{\prime }}  \notag \\
+e^{i\mathbf{k\cdot r}}e^{i\mathbf{k}^{\prime }\cdot \mathbf{r}^{\prime }}c_{%
\mathbf{k}}c_{\mathbf{k}^{\prime }}+(\mathbf{k,k}^{\prime }\rightarrow -%
\mathbf{k,-k}^{\prime })\}  \notag \\
+\frac{2i}{N^{2}}\sum_{\mathbf{k\in }A}\sin \left[ \mathbf{k}\cdot \left( 
\mathbf{r}^{\prime }-\mathbf{r}\right) \right] c_{-\mathbf{k}}c_{\mathbf{k}},
\end{gather}%
and then

\begin{eqnarray}
&&c_{\mathbf{r}}c_{\mathbf{r}^{\prime }}  \notag \\
&=&\frac{1}{N^{2}}\sum_{\mathbf{k,k}^{\prime }\mathbf{\in }A,\left(
k_{x}^{\prime }>k_{x}\right) }\{[e^{i\mathbf{k\cdot r}}e^{i\mathbf{k}%
^{\prime }\cdot \mathbf{r}^{\prime }}-(\mathbf{k\leftrightarrows k}^{\prime
})]c_{\mathbf{k}}c_{\mathbf{k}^{\prime }}  \notag \\
&&+[e^{-i\mathbf{k\cdot r}}e^{-i\mathbf{k}^{\prime }\cdot \mathbf{r}^{\prime
}}-(\mathbf{k\leftrightarrows k}^{\prime })]c_{-\mathbf{k}}c_{-\mathbf{k}%
^{\prime }}  \notag \\
&&+[e^{i\mathbf{k\cdot r}}e^{-i\mathbf{k}^{\prime }\cdot \mathbf{r}^{\prime
}}-(\mathbf{k\leftrightarrows -k}^{\prime })]c_{\mathbf{k}}c_{-\mathbf{k}%
^{\prime }}  \notag \\
&&+[e^{i\mathbf{k}^{\prime }\mathbf{\cdot r}}e^{-i\mathbf{k}\cdot \mathbf{r}%
^{\prime }}-(\mathbf{k\leftrightarrows -k}^{\prime })]c_{\mathbf{k}^{\prime
}}c_{-\mathbf{k}}\}  \notag \\
&&+\frac{2i}{N^{2}}\sum_{\mathbf{k\in }A}\sin \left[ \mathbf{k}\cdot \left( 
\mathbf{r}^{\prime }-\mathbf{r}\right) \right] c_{-\mathbf{k}}c_{\mathbf{k}}.
\end{eqnarray}%
Taking $\mathbf{r=}0$ and $\mathbf{r}^{\prime }=\mathbf{r}$, this relation
directly results in%
\begin{eqnarray}
g\left( r,\theta \right) &=&\left\vert \left\langle \text{G}\left( 0,\pi
/2\right) \right\vert c_{0}^{\dag }c_{\mathbf{r}}^{\dag }\left\vert \text{G}%
\left( 0,\pi /2\right) \right\rangle \right\vert  \notag \\
&=&\left\vert \frac{2i}{N^{2}}\sum_{\mathbf{k\in }A}\left\langle \text{G}%
\left( 0,\pi /2\right) \right\vert \sin \left( \mathbf{k}\cdot \mathbf{r}%
\right) c_{\mathbf{k}}^{\dag }c_{-\mathbf{k}}^{\dag }\left\vert \text{G}%
\left( 0,\pi /2\right) \right\rangle \right\vert  \notag \\
&=&\frac{2}{\pi ^{2}\left\vert y^{2}-x^{2}\right\vert }  \notag \\
&=&\frac{2}{\pi ^{2}r^{2}\left\vert \cos \left( 2\theta \right) \right\vert }%
,\left( x+y=\mathrm{odd}\right) .
\end{eqnarray}%
In addition, applying the relation to four operators 
\begin{eqnarray}
&&\left( c_{\mathbf{r}}c_{\mathbf{r}+\Delta \mathbf{r}}\right) ^{\dag }c_{%
\mathbf{r}^{\prime }}c_{\mathbf{r}^{\prime }+\Delta \mathbf{r}^{\prime }} 
\notag \\
&=&\frac{1}{N^{4}}\sum_{\mathbf{k}_{1}\mathbf{,k}_{2},\mathbf{k}_{3}\mathbf{%
,k}_{4}}e^{-i\mathbf{k}_{1}\cdot \left( \mathbf{r+\Delta r}\right) }e^{-i%
\mathbf{k}_{2}\mathbf{\cdot r}}  \notag \\
&&\times e^{i\mathbf{k}_{3}\mathbf{\cdot r}^{\prime }}e^{i\mathbf{k}%
_{4}\cdot \left( \mathbf{r}^{\prime }+\Delta \mathbf{r}^{\prime }\right) }c_{%
\mathbf{k}_{1}}^{\dag }c_{\mathbf{k}_{2}}^{\dag }c_{\mathbf{k}_{3}}c_{%
\mathbf{k}_{4}}.
\end{eqnarray}
The result

\begin{eqnarray}
&&\left\langle \text{G}\left( 0,\pi /2\right) \right\vert c_{\mathbf{k}%
_{1}}^{\dag }c_{\mathbf{k}_{2}}^{\dag }c_{\mathbf{k}_{3}}c_{\mathbf{k}%
_{4}}\left\vert \text{G}\left( 0,\pi /2\right) \right\rangle  \\
&=&\frac{1}{4}\left\{ 
\begin{array}{c}
\delta _{\mathbf{k}_{1},\mathbf{k}_{4}}\delta _{\mathbf{k}_{2},\mathbf{k}%
_{3}},\left( \left\vert k_{1x}\right\vert >\left\vert k_{2x}\right\vert
,\left\vert k_{4x}\right\vert >\left\vert k_{3x}\right\vert \right)  \\ 
-\delta _{\mathbf{k}_{3},\mathbf{k}_{1}}\delta _{\mathbf{k}_{2},\mathbf{k}%
_{4}},\left( \left\vert k_{1x}\right\vert <\left\vert k_{2x}\right\vert
,\left\vert k_{4x}\right\vert >\left\vert k_{3x}\right\vert \right)  \\ 
-\delta _{\mathbf{k}_{3},\mathbf{k}_{1}}\delta _{\mathbf{k}_{2},\mathbf{k}%
_{4}},\left( \left\vert k_{1x}\right\vert >\left\vert k_{2x}\right\vert
,\left\vert k_{4x}\right\vert <\left\vert k_{3x}\right\vert \right)  \\ 
\delta _{\mathbf{k}_{1},\mathbf{k}_{4}}\delta _{\mathbf{k}_{2},\mathbf{k}%
_{3}},\left( \left\vert k_{1x}\right\vert <\left\vert k_{2x}\right\vert
,\left\vert k_{4x}\right\vert <\left\vert k_{3x}\right\vert \right)  \\ 
2\delta _{\mathbf{k}_{3},-\mathbf{k}_{4}}\delta _{\mathbf{k}_{1},-\mathbf{k}%
_{2}}\delta _{\mathbf{k}_{2},\mathbf{k}_{3}}, \\ 
-2\delta _{\mathbf{k}_{3},-\mathbf{k}_{4}}\delta _{\mathbf{k}_{1},-\mathbf{k}%
_{2}}\delta _{\mathbf{k}_{2},-\mathbf{k}_{3}}, \\ 
\text{\textrm{S}}\left( \mathbf{k}_{3}\right) \text{\textrm{S}}\left( 
\mathbf{k}_{2}\right) \delta _{\mathbf{k}_{3},-\mathbf{k}_{4}}\delta _{%
\mathbf{k}_{1},-\mathbf{k}_{2}},\left( \mathbf{k}_{3}\neq \pm \mathbf{k}%
_{2}\right) 
\end{array}%
\right.   \notag
\end{eqnarray}%
where \textrm{S} is defined%
\begin{equation}
\text{\textrm{S}}\left( \mathbf{k}\right) =\mathrm{sign}\left( \sin
k_{x}-\sin k_{y}\right) ,
\end{equation}%
we have

\begin{eqnarray}
&&C_{\mathbf{RR}^{\prime }}=|\frac{1}{N^{4}}\sum_{\mathbf{k}_{2},\mathbf{k}%
_{3}\in A,\left( \mathbf{k}_{3}\neq \mathbf{k}_{2}\right) }\sin \left( 
\mathbf{k}_{2}\cdot \Delta \mathbf{r}\right) \sin \left( \mathbf{k}_{3}%
\mathbf{\cdot }\Delta \mathbf{r}^{\prime }\right)  \notag \\
&&+\frac{1}{4N^{4}}\sum_{\left\vert k_{1x}\right\vert \neq \left\vert
k_{2x}\right\vert }e^{i\left( \mathbf{k}_{1}+\mathbf{k}_{2}\right) \mathbf{%
\cdot }\left( \mathbf{r}^{\prime }\mathbf{-r}\right) }  \notag \\
&&\times \left[ e^{i\mathbf{k}_{1}\cdot \left( \Delta \mathbf{r}^{\prime
}-\Delta \mathbf{r}\right) }-e^{i\mathbf{k}_{2}\mathbf{\cdot }\Delta \mathbf{%
r}^{\prime }}e^{-i\mathbf{k}_{1}\cdot \Delta \mathbf{r}}\right] |,
\end{eqnarray}

where $\mathbf{R=r}+\Delta \mathbf{r}$\ and $\mathbf{R}^{\prime }\mathbf{=r}%
^{\prime }+\Delta \mathbf{r}^{\prime }$. In the thermodynamic limit, it can
be integrated as

\begin{eqnarray}
&&C_{\mathbf{RR}^{\prime }}=|\Gamma \left( \Delta \mathbf{r}^{\prime
}\right) \Gamma \left( \Delta \mathbf{r}\right)  \notag \\
&&+\frac{1}{4}\left( \delta _{\mathbf{r}^{\prime }\mathbf{,r}}\delta
_{\Delta \mathbf{r,\Delta r}^{\prime }}-\delta _{\mathbf{r}^{\prime }\mathbf{%
,r+\Delta r}}\delta _{\mathbf{r,r}^{\prime }+\Delta \mathbf{r}^{\prime
}}\right) |.
\end{eqnarray}

\end{document}